\documentclass[twocolumn,revtex4,apj,iop, twocolappendix, numberedappendix]{openjournal}
\usepackage{xcolor}
\usepackage{graphicx}
\usepackage{amsfonts}
\usepackage{amssymb, bm}
\usepackage{url}
\usepackage{amsmath}
\usepackage[varg]{txfonts}
\usepackage[normalem]{ulem}
\usepackage{fontawesome}
\usepackage[breaklinks,colorlinks,citecolor=blue,linkcolor=blue,urlcolor=blue]{hyperref}
\usepackage{booktabs}
 
\DeclareMathAlphabet{\mathbbold}{U}{bbold}{m}{n}

\renewcommand{\d}{\mathrm{d}}

\definecolor{MyB}{rgb}{0.1,0.1,1.0}

\setcitestyle{numbers,square}
\begin{document}

\title{Learning the averaged history of an inhomogeneous universe from its present day density field}

\author{Jonas Broe Bendtsen$^{\;a,b}$\footnote{jonasbb@student.matnat.uio.no}}
\author{Sofie Marie Koksbang$^{\;a}$ \vspace{0.15cm}\footnote{koksbang@cp3.sdu.dk}}

\affiliation{${}^a$ CP3-Origins, University of Southern Denmark, Campusvej 55, DK-5230 Odense M, Denmark}
\affiliation{${}^b$ Institute of Theoretical Astrophysics, University of Oslo, 0315 Oslo, Norway}

\begin{abstract}
In this work, we investigate whether machine learning can be used to infer averaged cosmological quantities directly from the present day matter density distribution. Using an implementation of the simplified silent universe framework, we generate a dataset consisting of 92610 independent relativistic, simplified cosmological simulations spanning a range of initial conditions and average cosmological parameters. We train a convolutional neural network to take the present-time matter density field as input and predict the averaged matter, cosmological constant, curvature, and kinematical backreaction density parameters as well as the Hubble parameter, at both initial and present time. The network achieves coefficients of determination exceeding 0.9 for all predicted quantities, with error distributions showing that only a small fraction of predictions reach percent-level errors or above. Although our use of the simplified silent universe approximation precludes direct application of the trained model to observational data, the results provide a proof-of-principle that neural networks can successfully recover averaged cosmological properties, including backreaction, at different epochs, simply from the late-time matter distribution. More broadly, our findings demonstrate that information about the averaged cosmological history of a universe is encoded in its present-time density field and can be extracted using machine learning techniques. This opens the possibility of applying similar approaches to more realistic cosmological simulations and observational probes such as N-body simulations and weak-lensing maps, ultimately providing a new avenue for constraining the large-scale evolution of the Universe.
\end{abstract}

\maketitle 
\tableofcontents

\section{Introduction}
The standard cosmological model describes the Universe as homogeneous and isotropic on sufficiently large scales, representing its evolution through the Friedmann--Lemaître--Robertson--Walker (FLRW) spacetimes. This approximation has proven remarkably successful. Nonetheless, the actual universe exhibits inhomogeneities in the form of e.g. galaxies, galaxy clusters, filament and voids. Since Einstein's field equations are nonlinear, averaging and time evolution is non-commutative in general relativity \cite{fluid1}. This means that the average/large-scale evolution of an inhomogeneous universe cannot in general be described by the dynamics of a single, global FLRW model. The deviation between FLRW evolution and the evolution of averaged inhomogeneous spacetimes is known as cosmic backreaction and is usually studied within Buchert's averaging scheme \cite{fluid1, fluid2}. This scheme employs volume-weighted scalar averages known to be related in simple manners to mean observations of redshift, the redshift-distance relation and cosmic chronometers \cite{syksy_light1, syksy_light2, selv1, selv2, selv_cc, asta_cc}. Within this averaging framework, volume-weighted spatial averages of the Raychaudhuri equation and Hamiltonian constraint lead to dynamical equations very similar to the Friedmann and acceleration equations of the FLRW spacetimes. Specifically, the obtained average dynamical equations are (considering, for simplicity, the irrotational dust limit with spatial hypersurfaces orthogonal to the dust flow)
\begin{align}
    3\frac{\dot{a}_\mathcal{D}^2}{a_\mathcal{D}^2} &= 3H_\mathcal{D}^2 = - \frac{1}{2} ^{(3)}\mathcal{R}_\mathcal{D} + 8\pi G\rho_\mathcal{D} + \Lambda - \frac{1}{2} Q_\mathcal{D}, \\
    3\frac{\ddot{a}_\mathcal{D}}{a_\mathcal{D}} &= Q_\mathcal{D} - 4\pi G\rho_\mathcal{D} + \Lambda .
\end{align}
Subscripts $\mathcal{D}$ indicate spatial volume-weighted averages defined as
\begin{align}
    s_{\mathcal{D}}:=\frac{\int_{\mathcal{D}}s\sqrt{|\det ^{(3)}g|}\d^3x}{\int_{\mathcal{D}}\sqrt{|\det ^{(3)}g|}\d^3x} = \frac{\int_{\mathcal{D}}s\sqrt{|\det ^{(3)}g|}\d^3x}{V_{\mathcal{D}}},
\end{align}
where $s$ is some scalar, and $\sqrt{|\det ^{(3)}g|}$ is the square root of the absolute value of the determinant of the spatial part of the metric tensor such that $V_\mathcal{D}$ is the proper volume of the spatial domain $\mathcal{D}$. The kinematical backreaction $Q_{\mathcal{D}}$ is given by $Q_\mathcal{D} = \frac{2}{3}\big((\Theta^2 )_\mathcal{D} - ( \Theta _\mathcal{D})^2\big)  - 2\sigma^2 _\mathcal{D}$, combining local expansion, $\Theta$, and the shear scalar $\sigma^2 = \frac{1}{2} \sigma_{\mu\nu}\sigma^{\mu\nu}$ of the fluid. The kinematical backreaction is a term arising when averaging the local equations, and does not appear in the Friedmann equation. The other backreaction contribution to the average evolution comes from the curvature term $^{(3)}\mathcal{R}_\mathcal{D}$. This average curvature can evolve differently than the curvature in FLRW models which must always scale inversely proportional to the squared scale factor. The averaged expansion rate $H_\mathcal{D}=\frac{\dot{a}_\mathcal{D}}{a_\mathcal{D}} = \frac{1}{3} \Theta_\mathcal{D}$ is the average of the expansion scalar in the considered averaging domain $\mathcal{D}$ and is equivalently the time derivative of the volume-scale factor, $a_\mathcal{D}$, defined as the cube root of the proper volume of the region, usually normalized to 1 at present time.
\newline\indent
Over the past decades, cosmic backreaction has attracted considerable attention as a possible contributor to the effective large-scale dynamics of the Universe, with some recent examples being \cite{recent_bc1, recent_bc2, recent_bc3, recent_bc4, recent_bc5, recent_bc6}.
\newline\indent
Despite extensive theoretical work, the observational detection and quantification of cosmic backreaction remains challenging. Direct observational constraints on backreaction have only recently become available through two different approaches. One approach \cite{recent_bc6} has relied on cosmicflows-4-based \cite{cosmicflows4} reconstructions of the local Universe, allowing estimates of large-scale curvature and averaged cosmological quantities. While promising, the analysis depends on reconstruction methods and modeling assumptions whose impact is difficult to fully quantify. A complementary observational approach \cite{asta_bc} employs cosmological distance measurements to directly constrain averaged curvature and backreaction terms. However, current constraints are weak \cite{addendum} and may exhibit degeneracies with other effects related to light propagation in an inhomogeneous universe such as the Dyer-Roeder effect \cite{DR}. Furthermore, the approach of \cite{asta_bc} does not constrain curvature and kinematical backreaction independently, but as the specific combination $\Omega_R + 3\Omega_Q$ (where density parameters are defined in the usual manner -- see e.g. \cite{addendum} for details). Although these recent developments represent important progress, independent methods for constraining cosmic backreaction are clearly still desirable. We here propose and explore the idea that deep neural networks can be useful for this endeavor.
\newline\indent
Machine learning techniques have emerged as powerful tools in cosmology. Neural networks and decision trees have demonstrated the ability to extract cosmological information such as the present-time density parameter $\Omega_{m,0}$, and $\sigma_8$ from various types of data sets. For instance, \cite{nnDM1, nnDM2, nnDM3} used neural networks to extract cosmological parameters such as $\Omega_{m,0}$ from the distribution of matter. In \cite{nnDM3}, the authors also demonstrated that the cosmological parameters can further be obtained through random forests trained on the non-linear matter power spectrum, and in \cite{nnpower1}, the cosmological parameters were extracted from dark matter halo densities instead. In \cite{nnlightcone}, the authors trained neural networks on lightcone data and applied it to SDSS data \cite{sdss} to extract $\Omega_{m,0}$ and $\sigma_8$. Light cone information was also used in \cite{nnkids}, where neural networks were trained to extract cosmological parameter information from weak lensing data from KiDS-450 \cite{kids}. Also \cite{nnlensing1, nnlensing2, nnlensing3} trained neural networks to constrain cosmological parameters with weak lensing data. Other studies use neural networks for learning about non-standard cosmology. Examples e.g. include \cite{chegeniClusternetsDeepLearning2024} where a neural network for detecting clustering dark matter was developed, and \cite{nine} where different models of modified gravity and massive neutrino cosmologies were classified using neural networks. Overall, these examples suggest that machine learning may be capable of identifying subtle signatures of cosmological phenomena that are otherwise difficult to isolate.
\newline\indent
Motivated by these advances, we investigate whether information related to cosmic backreaction is encoded in the present day matter distribution and can be extracted using neural networks. Specifically, we study whether a neural network can infer volume averaged cosmological quantities, including backreaction and curvature, from the late time density field alone. If successful, this would demonstrate that information related to the averaged expansion history of the Universe remains imprinted in the evolved matter distribution. The implications of such a result extend beyond the problem of cosmic backreaction itself by addressing the broader question of whether observations of the late universe can be used to recover information about the conditions and evolution of the Universe at earlier epochs. In this sense, the present day density field may be viewed as a fossil record of cosmic evolution, with machine learning providing a framework for decoding the information it contains. Establishing that such information is recoverable even in the presence of non-linear structure formation would open new possibilities for connecting late time observations to the physics governing the Universe across cosmic time.
\newline\indent
The goal of this work is not to derive new observational constraints on cosmic backreaction. Instead, we seek to establish the feasibility of a new methodology. Using ensembles of advanced toy-models introduced in section \ref{sec:data}, we train neural networks (see section \ref{sec:network}) to infer cosmological quantities, at both early and late times, directly from present day matter density maps. Our results, presented in section \ref{sec:results}, suggest that neural network-based analyses may eventually provide a complementary route for constraining cosmic backreaction once trained on realistic simulations.
\begin{table}
    \centering
    \caption{Parameters used to generate initial conditions with \texttt{CLASS}.}
    \small
    \label{tab:training_data}
    \begin{tabular}{llc}
        \toprule
        \textbf{Parameter} & \textbf{Value} \\
        \midrule
        Initial $z$ & 90\\
        Final $z$ & 0\\
        $n_s$ & 0.966 \\
        $\sigma_8$ & 0.81 \\
        $\Omega_{m,0}$ & $\{0.20,0.21,...,0.39,0.40\}$\\
        $\Omega_{\Lambda,0}$ & $\{0.60,0.61,...,0.79,0.80\}$ \\
        $h$ & $\{0.60,0.61,...,0.79,0.80\}$\\
        Number of cells & $64\times64\times64=262144$ \\
        Physical simulation size & $256\,\mathrm{Mpc}/h$ \\
        Version of each combination & 10 \\
        \midrule
        \textbf{Total number of simulations} & 92610\\
        \bottomrule
    \end{tabular}
\end{table}
\begin{figure*}
    \centering
    \includegraphics[width=0.9\linewidth]{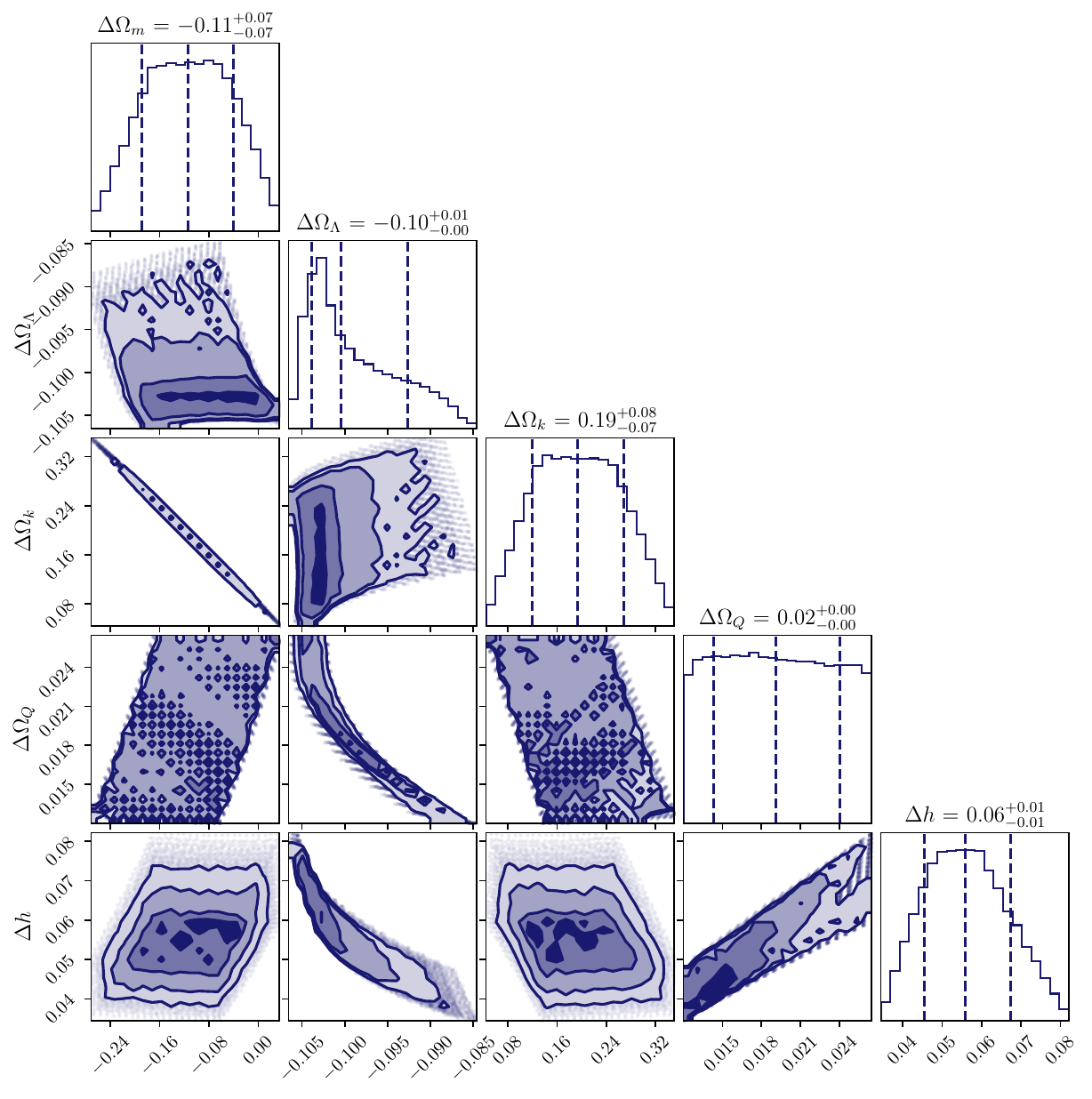}
    \caption{Corner plot showing relations between the cosmological parameters. We define $\Delta\Omega_x = \Omega_{x,0}^\mathcal{D}-\Omega_{x,0}^\mathrm{FLRW}$ (and similarly for $\delta h$)}
    \label{fig:corner}
\end{figure*}

\begin{table}
    \centering
    \caption{Architecture of our Convolutional Neural Network.}
    \small
    \label{tab:model_summary}
    \begin{tabular}{llc}
        \toprule
        \textbf{Layer type} & \textbf{Output shape} & \textbf{Parameters} \\
        \midrule
        Input & (64, 64, 64, 1) & 0 \\
        \midrule
        Conv3D (16 filters, $3\times3\times3$) & (64, 64, 64, 16) & 448 \\
        BatchNormalization & (64, 64, 64, 16) & 64 \\
        MaxPooling3D ($2\times2\times2$) & (32, 32, 32, 16) & 0 \\
        \midrule
        Conv3D (32 filters, $3\times3\times3$) & (32, 32, 32, 32) & 13,856 \\
        BatchNormalization & (32, 32, 32, 32) & 128 \\
        MaxPooling3D ($2\times2\times2$) & (16, 16, 16, 32) & 0 \\
        \midrule
        Conv3D (64 filters, $3\times3\times3$) & (16, 16, 16, 64) & 55,360 \\
        BatchNormalization & (16, 16, 16, 64) & 256 \\
        MaxPooling3D ($2\times2\times2$) & (8, 8, 8, 64) & 0 \\
        \midrule
        Conv3D (128 filters, $3\times3\times3$) & (8, 8, 8, 128) & 221,312 \\
        BatchNormalization & (8, 8, 8, 128) & 512 \\
        MaxPooling3D ($2\times2\times2$) & (4, 4, 4, 128) & 0 \\
        \midrule
        GlobalAveragePooling3D & (128) & 0 \\
        Dense (128 units) & (128) & 16,512 \\
        Dropout (rate=0.3) & (128) & 0 \\
        Dense (10 units) & (10) & 1,290 \\
        \midrule
        \multicolumn{2}{l}{\textbf{Total parameters:}} & \textbf{309,738} \\
        \bottomrule
    \end{tabular}
\end{table}
\begin{figure}
    \centering
    \includegraphics[width=\linewidth]{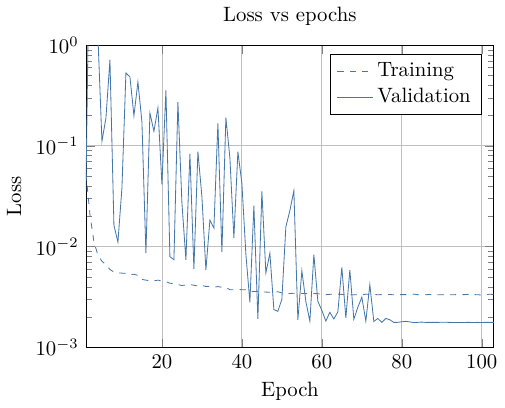}
    \caption{Training loss (solid line) and validation loss (dashed line) of the CNN, displayed using a logarithmic y-axis. The plots show the loss decreasing with an increasing number of training epochs.}
    \label{fig:loss}
\end{figure}
\section{Training data}\label{sec:data}
A major challenge in studies using supervised machine learning is the generation of suitable training data. This challenge becomes even more significant when our data is to be based on general relativity where running numerical codes typically require significant computational resources. Several tools are available today for simulating relativistic spacetimes, where \texttt{gevolution} \cite{gevolution} and a scheme \cite{ET_hayley} based on the \texttt{Einstein Toolkit} \cite{toolkit1, toolkit2} seem to be the most used for purely cosmological purposes. The former has recently been upgraded to run on GPUs \cite{gpu}, making it more realistic to obtain large catalogs of simulation data necessary for training neural networks. However, neither of these simulation frameworks have been found to lead to significant backreaction when averaging over the entire simulation domain \cite{ETbc1, ETbc2, gevolutionbc}. This may in principle be a genuine feature of cosmological spacetimes with realistic initial conditions set in agreement with the Cosmic Microwave Background (CMB). However, there is reason to believe that it may instead be an artifact and that backreaction effects are suppressed in cosmological simulations with periodic boundary conditions (see e.g. \cite{boundary}). In principle, this hypothesis could be tested by making a very large simulation and studying the evolution only in a region around the center of the simulation box, causally disconnected from the boundaries, to see if this central region developed significant backreaction. Unfortunately, this approach does not seem to be possible with existing simulations. For \texttt{gevolution} simulations, the method is futile since the code solves constraint equations which are elliptic and determined on the entire spatial hypersurfaces of the simulations at once. Therefore, all spatial points in a single time stamp of the simulation are instantaneously connected with each other, and the central part of the simulation will at all times be affected by the periodic boundary conditions. The cosmological simulations of \cite{ET_hayley} do not have this problem as they solve the fully hyperbolic Einstein field equations. However, in this case, making a large simulation to study only the central part may still be futile since these simulations are only mildly non-linear. With backreaction being a non-linear relativistic effect, we thus expect that backreaction cannot yet be realistically quantified with the simulations of \cite{ET_hayley}.
\newline\indent
Overall, it is currently unclear to what extent we can consider these simulations realistic in terms of backreaction, and hence also to what extent we can trust their apparent results indicating that backreaction is small on global scales. We will therefore here take the agnostic point of view and consider the size and relevance of backreaction in the real universe an open question.
\newline\newline
Due to the uncertainty of the situation discussed above combined with considerations of computational resources, we here choose to use simplified simulations known to develop clear backreaction signatures on global scales. Specifically, we will use the simplified silent universe scheme introduced in \cite{simsilun} as detailed in the subsection below. This lets us compile a large training data set that exhibits non-trivial average cosmic histories including various levels of backreaction. This is sufficient for our proof-of-principle study, but we stress that the training data is not realistic enough for it to be sensible to apply the resulting trained network to real data.

\subsection{Simulation data with the simplified silent universe approximation}
The Silent universe approximation is formulated in the the 1+3 formalism of general relativity \cite{13_oldie, 13_ellis}, and corresponds to the special situation where individual world lines evolve independently and thus represent solutions to Einstein's equations without e.g. gravitational waves and energy flux. Within this approximation, the evolution of the spatial hypersurfaces orthogonal to the cosmic fluid are solutions to the coupled ordinary differential equations \cite{simsilun}
\begin{align}
    \dot{\rho} &= -\rho \Theta\,,\label{eq:silent_rho}\\
    \dot{\Theta} &= -\frac{1}{3}\Theta^2 - 4\pi G \rho - 6 \Sigma^2 + \Lambda \,,\label{eq:silent_theta}\\
    \dot{\Sigma} &= -\frac{2}{3}\Theta \Sigma + \Sigma^2 - \mathcal{W} \,,\label{eq:silent_sigma}\\
    \dot{\mathcal{W}} &= -\Theta \mathcal{W} - 4\pi G \rho \Sigma - 3 \Sigma \mathcal{W} \,,\label{eq:silent_W} \\
    \dot{V} &= \Theta V \label{eq:silent_V}  \,,
\end{align}
where $\rho$ is the density, $\Theta$ is the expansion rate, $\Sigma$ the shear, $\mathcal{W}$ is the scalar representing the electric part of the Weyl tensor (as discussed in \cite{simsilun}, in a silent universe, the shear and electric Weyl tensor can be diagonalized, and each has only a single independent component). $V$ is the proper volume of a single simulation cell, defined up to an arbitrary scaling. The proper volume is used for calculating averages in the simulation.
\newline\indent
In \cite{simsilun}, an approximation scheme for setting up initial conditions for this set of equations was introduced. The resulting cosmological model is the {\em simplified silent universe}. The corresponding initial conditions can be summarized as
\begin{align}\label{eq:ic}
    \rho_i &= \bar{\rho}(1 + \delta_i) \,, \\
    \Theta_i &= \bar{\Theta}\bigg(1 - \frac{1}{3}\delta_i\bigg)\,, \\
    \Sigma_i &= \frac{1}{9}\bar{\Theta} \delta_i\,, \\
    \mathcal{W}_i &= -\frac{4\pi G}{3}\bar{\rho} \delta_i\,,\\
    V_i &= \frac{1}{1+\delta_i} \,.
\end{align}
At initial time, the background density and expansion rate are assumed to be those of an FLRW dust universe with possible non-zero curvature. At initial time we therefore have
\begin{align}
    \bar{\rho} &= \frac{3 H_0^2 \Omega_{m,0}}{8\pi G} (1 + z_i)^3, \\
    \bar{\Theta} &= 3 H_0 \sqrt{\Omega_{m,0} (1 + z_i)^3 + \Omega_{\Lambda,0} + \Omega_{k,0} (1 + z_i)^2}.´,
\end{align}
where the cosmological parameters $H_0, \Omega_{m,0}, \Omega_{k,0}$ and $\Omega_{\Lambda,0}$ will be specified later.
\newline\indent
We have written our own Julia \cite{bezansonJuliaFreshApproach2017} implementation of the simplified silent universe approximation described above.\footnote{The code is publicly available at \href{https://github.com/jbb42/simsilun_backreaction/}{\faGithub/jbb42/simsilun\_backreaction}.} Our code testing included e.g. verifying that it can exactly reproduce the evolution of Lemaitre-Tolman-Bondi (LTB) models \cite{ltb1, ltb2, ltb3}, which are exact silent solutions to Einstein's equations. We also studied the error obtained by applying the ``simplified'' assumption when generating initial conditions for the LTB reproductions. Although this error can become significant in some regions, this appears to mainly happen when structures cover multiples simulation cells which is clearly the case for the reproduced LTB models, but not something we would expect to be relevant for setting initial conditions based on perturbation theory as we do when setting up the initial conditions for our training data simulations. Details on the LTB tests can be found in appendix \ref{app:LTB}.
\newline\newline
A large ensemble of training data is needed to train the machine learning model. The parameters used to generate this ensemble are shown in table \ref{tab:training_data}. \texttt{CLASS} \cite{class1, class2} is used to set up the initial conditions at $z=90$ by generating matter power spectra using the parameters in table \ref{tab:training_data}. From each unique power spectrum, ten different Gaussian random fields are generated. These are used to obtain the density contrast, from which the initial conditions of the five cosmological parameters in eq. \ref{eq:ic} are found. Note that the table shows the values of $\Omega_{m,0}$ and $\Omega_{\Lambda,0}$ provided to \texttt{CLASS}. Since \texttt{CLASS} assumes FLRW evolution, the actual values of $\Omega_{m,0}$ and $\Omega_{\Lambda,0}$ obtained after averaging the simulations at present time will in general deviate from these values because of the non-negligible backreaction. $h_i^\mathcal{D}$ and $h_0^\mathcal{D}$ are defined as the average dimensionless Hubble parameter at initial and present time, respectively. In a homogeneous universe, $h_0^\mathcal{D}$ would thus simply be equal to the ordinary reduced Hubble parameter, $h$.
\newline\newline
The initial universe slices are evolved forward in time by using the silent universe equations \ref{eq:silent_rho}-\ref{eq:silent_V} until the timestep at which the background FLRW model used when setting up initial conditions reaches present time. The time step at which this happens is considered present time of the simulation. This stopping criterion differs from the approach in \cite{simsilun}, where the evolution time is calculated independently for each cell based on the corresponding FLRW model. We choose the slightly different stopping criterion as it seems to be closer in spirit to how traditional relativistic N-body simulations are run; these also stop at a specific coordinate time, but due to fluctuations in the metric tensor, each cell of the simulation will have slightly different proper time, exactly as in our simplified silent simulations.
\newline\newline
The corner plot in figure \ref{fig:corner} demonstrates how the 92610 simulations lead to average evolutions deviating from those of the FLRW models used for setting initial conditions. The corner plots show $\Delta\Omega_x := \Omega_{x,0}^\mathcal{D}-\Omega_{x,0}^\mathrm{FLRW}$ and $\delta h $ defined equivalently. The figure shows that cosmic backreaction emerges primarily as the curvature parameter $\Omega_{k,0}^\mathcal{D}$, which is systematically larger than its FLRW counterpart. This is in agreement with the results \cite{simsilun} obtained with the original implementation of the simplified silent universe approximation in the code \texttt{simsilun} as well as with the observational constraints of \cite{recent_bc6}.
\newline\indent
In the corner plot, we see that $\Omega_{m,0}^\mathcal{D}$ and $\Omega_{\Lambda,0}^\mathcal{D}$ are both lower than their FLRW counterparts. This is evident from the strong negative correlation between $\Delta\Omega_m$ and $\Delta\Omega_k$, which forms a narrow line in the corner plot. The effect of backreaction is further demonstrated by a strictly non-zero $\Delta\Omega_Q$ and the dimensionless Hubble constant $h_0^\mathcal{D}$ being larger than in the FLRW case. We also observe this as a positive correlation between $\Delta\Omega_Q$ and $\Delta h$. 
\begin{figure*}
    \centering
    \includegraphics[width=0.33\textwidth]{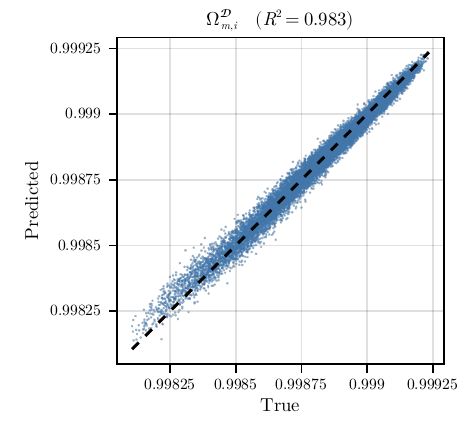}\hfill
    \includegraphics[width=0.33\textwidth]{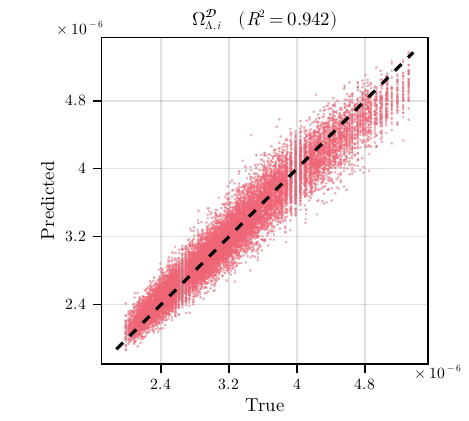}\hfill
    \includegraphics[width=0.33\textwidth]{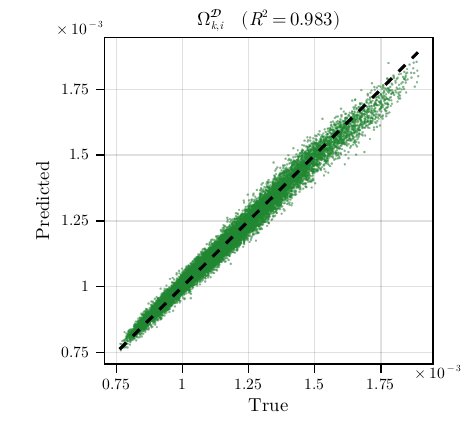}\\
    \includegraphics[width=0.33\textwidth]{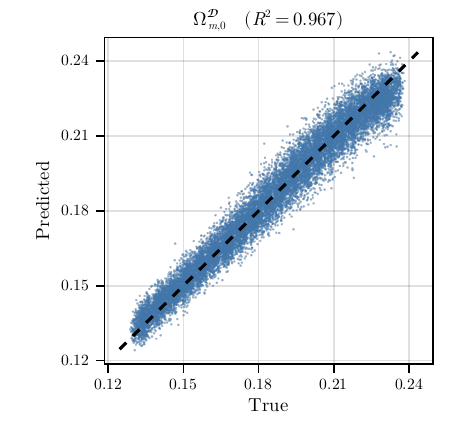}\hfill
    \includegraphics[width=0.33\textwidth]{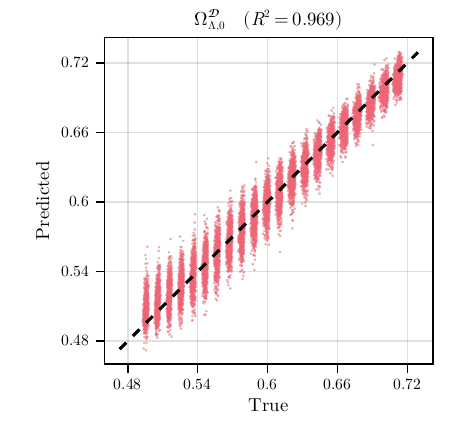}\hfill
    \includegraphics[width=0.33\textwidth]{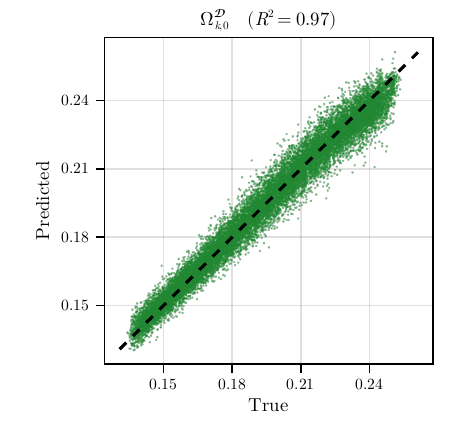}\\
    \includegraphics[width=0.33\textwidth]{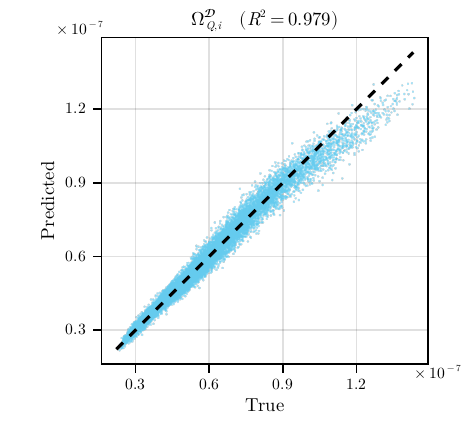}
    \includegraphics[width=0.33\textwidth]{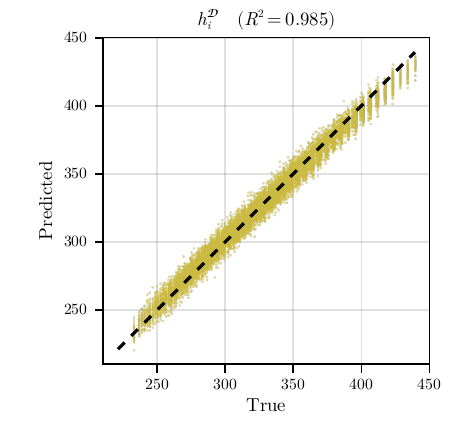}\hfill\\
    \includegraphics[width=0.33\textwidth]{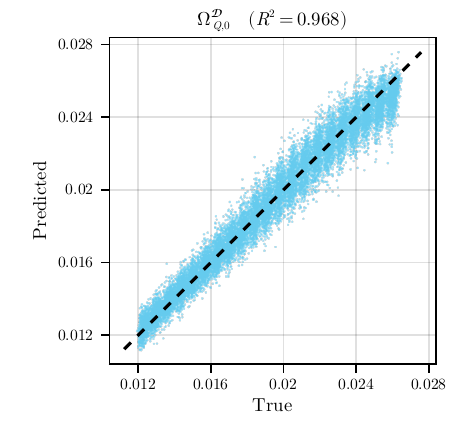}
    \includegraphics[width=0.33\textwidth]{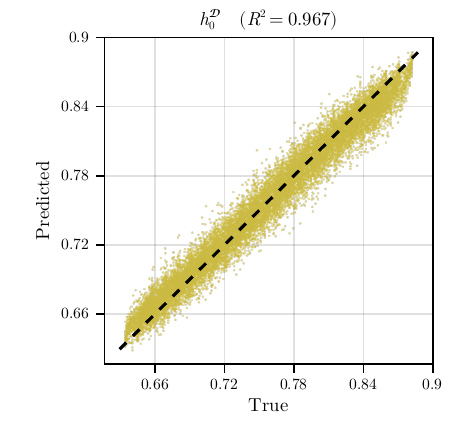}\hfill\\
    
    \caption{Comparison of true and predicted values, showing the initial and present-time $\Omega_m$ (blue points), $\Omega_\Lambda$ (pink points), $\Omega_k$ (green points), $\Omega_Q$ (cyan points), and the Hubble parameter $h$ (yellow points). The black dashed lines indicate where true and predicted values coincide.}
    \label{fig:cnn_all64}
\end{figure*}

\begin{figure*}
    \centering
    \includegraphics[width=0.31\textwidth]{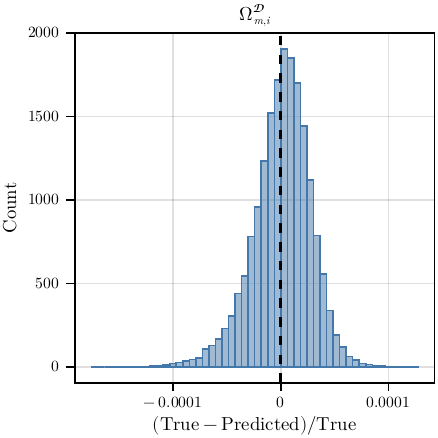}\hfill
    \includegraphics[width=0.31\textwidth]{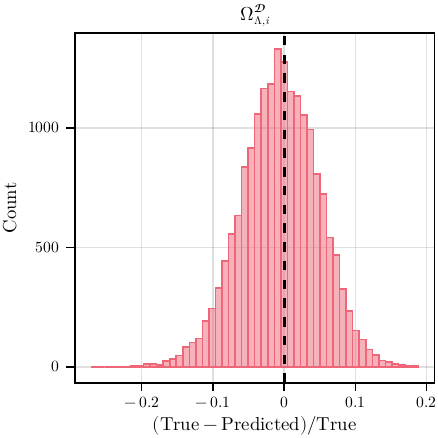}\hfill
    \includegraphics[width=0.31\textwidth]{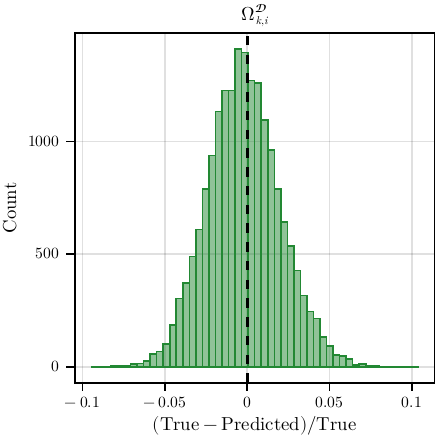}\\
    \includegraphics[width=0.31\textwidth]{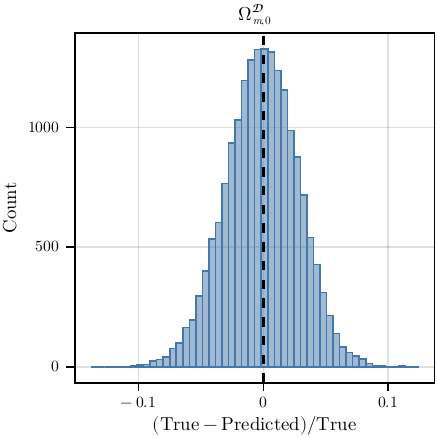}\hfill
    \includegraphics[width=0.31\textwidth]{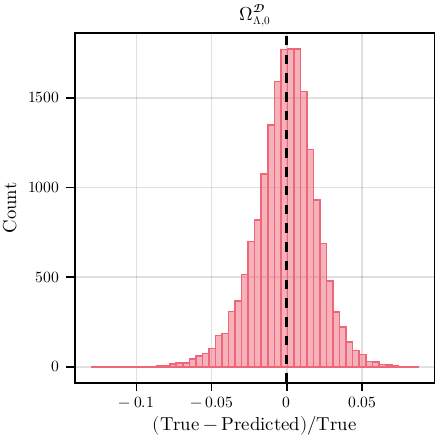}\hfill
    \includegraphics[width=0.31\textwidth]{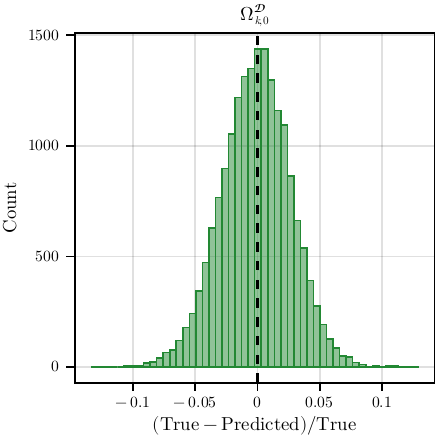}\\
    \includegraphics[width=0.31\textwidth]{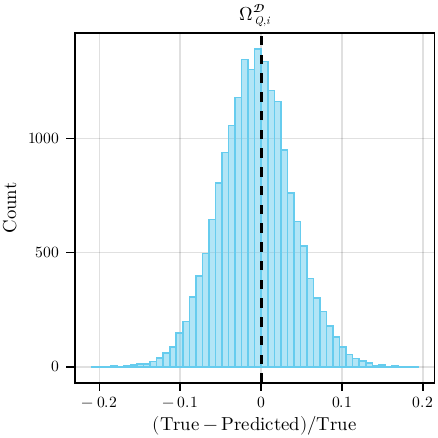}
    \includegraphics[width=0.31\textwidth]{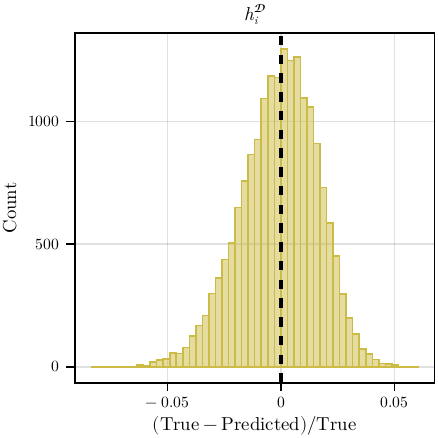}\hfill\\
    \includegraphics[width=0.31\textwidth]{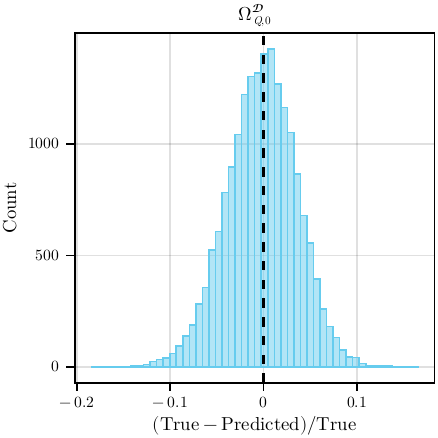}
    \includegraphics[width=0.31\textwidth]{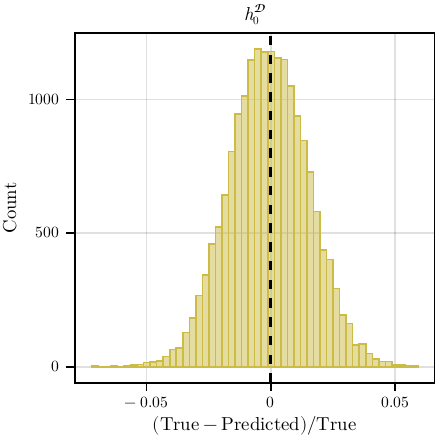}\hfill\\

    \caption{Histograms of the initial and final relative errors of the predicted parameters $\Omega_m$ (blue), $\Omega_\Lambda$ (pink), $\Omega_k$ (green), $\Omega_Q$ (cyan), and $h$ (yellow).}
    \label{fig:cnn_rel_err_hist}
\end{figure*}

\section{Neural Network Architecture}\label{sec:network}
A Convolutional Neural Network (CNN) \cite{osheaIntroductionConvolutionalNeural2015} is used to infer the initial and final/present-time density and Hubble parameters from the present-day density field. The architecture of the CNN is summarized in table \ref{tab:model_summary}. It is implemented in \texttt{TensorFlow}~\cite{abadiTensorFlowSystemLargescale2016} using the \texttt{Keras} interface~\cite{cholletKerasPythonDeep2018}, and was initially inspired by \cite{chegeniClusternetsDeepLearning2024}. Indeed, we initially worked with the architecture presented in \cite{chegeniClusternetsDeepLearning2024} but found it necessary to heavily modify the architecture before we obtained satisfactory results. For instance, we decreased the size of the kernel in the convolutional layers and introduced MaxPooling instead of a stride of 2 to downscale the data. 
\newline\newline
To train the network, the simulated data was split into 60\% training data, 20\% validation data, and 20\% test data. The training, validation, and test datasets contain equal shares of the different combinations of cosmological parameters to avoid bias. The training data is normalized to prevent the network from omitting parameters such as $\Omega_{Q,i}$ due to their small magnitudes. The input density map is normalized using Z-score normalization, while each cosmological parameter is normalized using min-max scaling between zero and one, thus ensuring that they are all weighted equally. The data is then split into batches consisting of 128 simulated universes each.

The network consists of four blocks of convolutional layers which applies the convolution filters, takes the Batch Normalization and downscales the data through MaxPooling. These are followed by a Global Average Pooling and a dense layer connected to the output layer with a dropout rate of 0.3. This structure aims to decrease overfitting while giving the network room to learn from the input data. The network is trained for up to 250 epochs, with an early stopping criterion implemented to halt the training and restore the optimal weights when the validation loss fails to improve for 20 successive epochs.

To optimize the network, we used \texttt{AdamW}~\cite{loshchilovDecoupledWeightDecay2019}, which is a modified version of the traditional \texttt{Adam} optimizer which incorporates weight decay into the optimization step. The weight decay coefficient is set to $0.002$ and the initial learning rate is set to $0.002$, with a dynamic learning rate schedule that reduces the learning rate by a factor of $0.5$ after five epochs with no decline in the validation loss. This reduction will happen until the learning rate reaches a minimum of $10^{-7}$. These parameters were obtained by manually tuning them until accurate reproduction of validation data was achieved.

We experimented with variations of this architecture, e.g. by adding a physics-inspired layer before the final output layer, which calculated the density parameters $\Omega_{x,i}$ and $\Omega_{x,0}$ from the inferred curvature, density, kinematical backreaction, and cosmological constant, requiring them to sum to 1. This produced significantly worse results than letting the network predict the final parameters independently, and was thus omitted in the final version of the network.

\section{Results}\label{sec:results}
The CNN described above was trained using 60\% of the simulated universes to predict the initial and final density parameters and Hubble parameters from the final density map. Figure \ref{fig:loss} shows a learning curve for our training of the network. As seen, the optimizer decreases the training and validation loss the first $\sim 100$ epochs after which the early stopping criterion sets in and restores the weights and biases from the epoch with the lowest validation loss. We see that the loss curves flatten out around epoch 80, indicating that the model has been trained for an optimal number of epochs already at that time.
\newline\indent
After training the network, we applied it to the test data. The results from this are shown in figure \ref{fig:cnn_all64}. The plots in this figure depict the predicted cosmological parameters of the test data against their true values, with the plots showing results for both initial and final parameter values. The comparisons show that the parameters are overall accurately predicted by the true values, with the coefficient of determination, $R^2$, ranging from $0.942$ to $0.985$. Despite being trained on present time data, the network is generally slightly better at predicting the initial parameters that the present time parameters. The only exception is $\Omega_{\Lambda}$, where the prediction of the initial values lead to $R^2 = 0.942$ while the present-time predictions achieve $R^2 = 0.969$. We also note that the density parameter of the cosmological constant is the one which is predicted with lowest $R^2$. Furthermore, especially its present-time plot in \ref{fig:cnn_all64} shows clear vertical lines. These are due to the discrete set of input $\Omega_\Lambda$-values used in \texttt{CLASS} and may suggest that a denser set of input values could improve the model.
\newline\indent
We also note that the network exhibits a general, slight tendency to under-predict the largest values of $\Omega_{Q,i}$. A similar feature is seen for $\Omega_{Q,0}$, $\Omega_{m,0}$, $\Omega_{\Lambda,i}$, $\Omega_{k,i}$ and $\Omega_{k,0}$, although much less prominently. Overall, this suggests that the model is somewhat conservative in the extreme upper part of the considered parameter space. 
\newline\newline
To further quantify the accuracy of the predictions, we have plotted the relative error as histograms in figure \ref{fig:cnn_rel_err_hist}. The histograms are roughly centered around zero, with fairly long but low tails. Focusing on the tails, we again see that initial values are more accurately reproduced than the present time parameter values, with the one clear exception being $\Omega_{\Lambda}$. For most of the parameters, the tails reach a maximum of about $\pm0.1-0.2$. The most significant exception is $\Omega_{m,i}$ which has a tail that only reaches the order $\sim 0.0001$.

\section{Summary and conclusion}\label{sec:conclusion}
The simplified silent universe simulations of \cite{simsilun} demonstrated that backreaction can be obtained from simplified silent universes. Using our own implementation of the simplified silent universe approximation, we generated training data corresponding to 92610 simulations with 20 \% of the data used as validation data, 20 \% retained for testing, and the remaining 60 \% constituting our training data. Our neural network takes the present-time matter density field as input and outputs the corresponding average density parameters of matter, the cosmological constant, curvature and kinematical backreaction at both initial and present time. The results are overall promising, with $R^2$ above 0.9 in all cases, and with error distributions showing that only small fractions of the predicted results for test data are of or above the percent level.
\newline\indent
Since our results are based on simplified simulations, they must be considered proof-of-principle and our trained model cannot sensibly be applied to real data. Nonetheless, as proof-of-principle, our study clearly demonstrates that with realistic training data, a network similar to ours should be able to output accurate values for averaged cosmological quantities.
\newline\indent
Although our motivation was to establish whether a neural network can be used to quantify cosmic backreaction in the Universe, our network demonstrates significantly more than that. In particular, our neural network demonstrates that it is possible to design neural networks that can output the average cosmological parameters at different times, from only the present time density map as input. In other words, our CNN demonstrates that neural networks can output the averaged history of the Universe, with only the present-time density distribution as input. It would be very interesting to expand this result to networks using more advanced simulation data as input, such as density maps from N-body simulations or weak lensing maps. We expect that the inference would become significantly more difficult in these cases since the simulations do so, but we see no reason why it should not be possible to obtain results similar to those we present here. We thus anticipate that the main obstacle would be the compilation of the dataset.

\acknowledgments
This project was funded by Villum Fonden, grant VIL53032 (PI: SMK).\\
{\bf Author contribution statement:} The presented work is based on results obtained by JBB as part of his master's thesis project conducted under the supervision of SMK who developed the broader research program and conceptual framework underlying this work. JBB carried out the numerical work including CNN design, under the guidance of SMK. The writing of the manuscript was a joint effort.\\
{\bf AI declaration:} Suggestions from large language models (LLMs) were used to optimize the code written for this project. No LLM suggestions were implemented without careful consideration by JBB.
\clearpage

\bibliography{references}

\appendix
\section{Simulation tests using LTB models}\label{app:LTB}
\begin{figure*}
    \centering
    \includegraphics[width=0.9\linewidth]{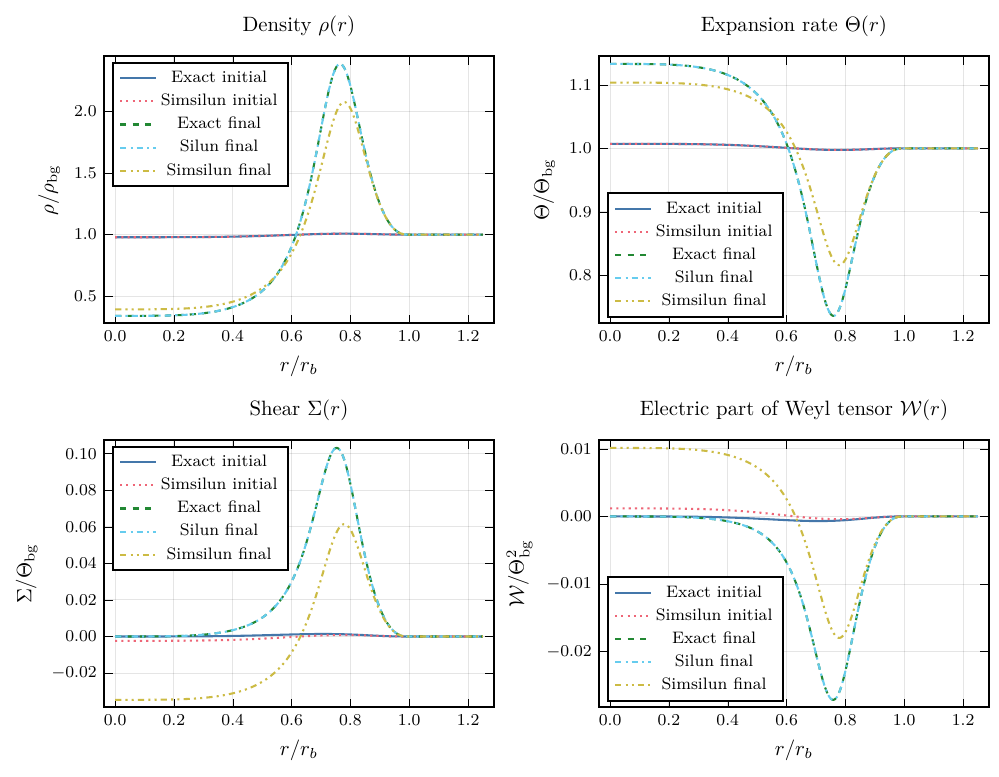}
    \caption{In reading order, the plots show the density $\rho(r)$ normalized to the background density $\rho_\mathrm{bg}$, the expansion rate $\Theta(r)$ normalized to the background expansion rate $\Theta_\mathrm{bg}$, the shear $\Sigma(r)$ which is also normalized to the background expansion rate $\Theta_\mathrm{bg}$, and the Weyl curvature $\mathcal{W}(r)$ normalized to the background expansion rate squared, $\Theta_\mathrm{bg}^2$. The plots depict the initial values from the exact solution (solid blue lines), the perturbative initial conditions (orange dotted lines), the exact final values (dashed green lines), the (overlapping) silent results (cyan dot-dashed lines), and the simplified silent simulation results (yellow dot-dot-dashed lines).}
    \label{fig:LTB}
\end{figure*}

To check our implementation of the simplified silent universe approximation, we test our simulation code's ability to reproduce a specific LTB model. The LTB model is itself silent, and should thus undergo the same evolution as a silent universe simulation as long as initial conditions correspond exactly to the LTB model. Any deviation between the exact LTB evolution and the simulated evolution would thus indicate bugs/errors in our simulation code. However, the simplified initial conditions do  not coincide with LTB models, even at early time. Comparing exact LTB results with the results obtained from our silent simulation with simplified initial condition therefore lets us assess the implications of introducing this approximation.
\newline\newline
The line element of LTB models can be written as
\begin{equation}\label{eq:LTBmetric}
    \d s^2 = -\d t^2 + \frac{A'(t, r)^2}{1-k(r)} \d r^2 + A(t, r)^2 \d \Omega^2 \,,
\end{equation}
where a prime denotes radial derivatives. The metric function $A(t,r)$ depends on both the position and time coordinate, and $k(r)$ is a curvature function depending only on the radial position. We use
\begin{equation}\label{eq:k_ltb}
    k(r) = \begin{cases}
        -r^2 k_\mathrm{max}\big((r/r_b)^n - 1\big)^m & r\leq r_b \\
        0 & r > r_b \,,
    \end{cases}
\end{equation}
with $k_\mathrm{max}=5.4\times10^{-8}$, and find $A(t,r)$ as the solution to the following differential equation:
\begin{equation}\label{eq:Adot}
    \dot{A}(t,r)^2 = -k(r) + \frac{2M(r)}{A(t,r)} + \frac{\Lambda}{3}A(t,r)^2 \,.,
\end{equation}
where a dot denotes time derivative. We solve this equation with initial conditions given by $A(t_i,r) = a_i r$, where $a$ is the scale factor of the $\Lambda$CDM model with $\Omega_{m,0} = 0.3$ and $H_0 = 70$km/s/Mpc, and a subscript $i$ indicates evaluation at initial time which we choose to be the coordinate time where $z = 1200$. For coordinate values $r>r_b$, the LTB model reduces exactly to the $\Lambda$CDM model defined above. We will refer to this as the background of the LTB model.
\newline\indent
The function $M(r)$ can be related to $k(r)$ through the relation \cite{strongtoweak}
\begin{equation}\label{eq:M_ltb}
    M(r) = \frac{4\pi G}{3} \rho_{\mathrm{FLRW},i} a_i^3  r^3 \Bigg(1 + \frac{3}{5}\frac{k(r)}{(a_i H_i r)^2}\Bigg) \,,
\end{equation}
valid at early times.
\newline\newline
When comparing results obtained from the simulations with the exact LTB evolution, we use the four quantities
\begin{align}
    \rho &= \rho = \frac{2M'}{\kappa A^2 A'}\,,\\
    \Theta &= \frac{\dot{A}'}{A'} + 2\frac{\dot{A}}{A} \,,\\
    \Sigma &= \frac{1}{9} \bigg( \frac{\dot{A}'}{A'} - \frac{\dot{A}}{A} \bigg)^2\,, \\
    \mathcal{W} &= \frac{M}{A^3}-\frac{M'}{3A^2A'} \,.
\end{align}
We plot $\rho$, $\Theta$, $\Sigma$, and $\mathcal{W}$ at coordinate times corresponding to $z = 90$ and present time of the background model. The plots are shown in figure \ref{fig:LTB} and include i) the exact LTB quantities, ii) the quantities evolved using the silent simulation but with exact LTB initial conditions set at $z = 90$, and iii) simplified, silent simulation results, where initial conditions for the simulation (set at $z = 90$) are set using the simplified scheme defined in equation \ref{eq:ic}. As seen, using the exact initial conditions of the LTB model in the simulation yields simulation results that closely reproduce the exact solution, indicating that our simulation code does not contain any severe errors/bugs. However, the silent initial conditions for the shear and Weyl curvature deviate significantly from the exact LTB values at $z = 90$. This leads $\rho$ and $\Theta$ to also deviate noticeably from the exact values at present-time, especially around the density peak. The final values of $\Sigma$ and $\mathcal{W}$ also differ significantly from the exact values. We expect that these discrepancies are due to the fact that the exact initial conditions for $\Sigma$ and $\mathcal{W}$ depend on the spatial gradients of $\delta$, whereas the simplified initial conditions depend only on the local value of $\delta$. Consequently, the perturbative and exact initial conditions differ significantly in regions with a non-zero density contrast and small density gradients, such as the void in the LTB model. The difference between the exact solution and the results obtained from the simplified silent simulations can therefore be attributed to inaccurate initial conditions for the shear and the Weyl curvature, and is mainly an issue when considering structures spanning many simulation cells.

\end{document}